\RequirePackage{fix-cm}
\documentclass[smallextended]{svjour3}       
\smartqed  
\usepackage{graphicx}
\usepackage{url}

\def\gtsim {>\kern-1.2em\lower1.1ex\hbox{$\sim$}~}   
\def\ltsim {<\kern-1.2em\lower1.1ex\hbox{$\sim$}~}   
\def \apj {Astrophysical Journal}
\def \apjl {Astrophysical Journal, Letter}
\def \apjs {Astrophysical Journal, Suppl}
\def \aj  {Astronomical Journal}
\def \aap {Astronomy \& Astrophysics} 
\def \mnras {Monthly Notices of the Royal Astronomical Society}

\def \nat {Nature}

\journalname{Experimental Astronomy}
\begin{document}

\title{The origin of elements: the need for UV spectra
}


\author{Chiaki Kobayashi
}


\institute{
Centre for Astrophysics Research,
Department of Physics, Astronomy and Mathematics
University of Hertfordshire,
College Lane, Hatfield  AL10 9AB, UK
              \email{c.kobayashi@herts.ac.uk}           
}

\date{Received: 29 January 2022 / Accepted: 9 July 2022}

\maketitle

\begin{abstract}
Thanks to the long-term collaborations between nuclear and astrophysics, we have good understanding on stellar nucleosynthesis, except for the elements around Ti and some neutron-capture elements.
From the comparison between observations and Galactic chemical evolution models, it is necessary to have the rapid neutron-capture process associated with core-collapse supernovae, although the explosion mechanism is unknown. 
The impact of rotating massive stars is also shown in this paper.
Many of the key elements can be exclusively obtained in the UV, and therefore without UV spectra it would not be possible to fully understand the origin of elements in the universe.

\keywords{stellar abundances \and  nucleosynthesis \and supernovae \and Milky Way Galaxy}
\end{abstract}

\section{Introduction}
\label{intro}
Explaining the origin of the elements is one of the scientific triumphs linking nuclear physics with astrophysics. As Fred Hoyle predicted, carbon and heavier elements (`metals' in astrophysics) were not produced during the Big Bang but instead created inside stars. So-called $\alpha$ elements (O, Ne, Mg, Si, S, Ar, and Ca) are mainly produced by core-collapse supernovae, while iron-peak elements (Cr, Mn, Fe, and Ni) are more produced by thermonuclear explosions, observed as Type Ia supernovae (SNe Ia, \cite{kob20sr}, hereafter K20). The production depends on the mass of white dwarf progenitors, and a large fraction of SNe Ia should come from near-Chandrasekhar (Ch) mass explosions (see \cite{kob20ia} for constraining the relative contribution between near-Ch and sub-Ch mass SNe Ia).
Among core-collapse supernovae, hypernovae produce a significant amount of Fe as well as Co and Zn, and a significant fraction of massive stars ($\gtsim 20M_\odot$) should explode as hypernovae in order to explain the Galactic chemical evolution (GCE, \cite{kob06}).

Heavier elements are produced by neutron-capture processes. The slow neutron-capture process (s-process) occurs in asymptotic giant branch (AGB) stars \cite{kar16}, while the astronomical sites of rapid neutron-capture process (r-process) have been debated. The possible sites are neutron-star (NS) mergers \cite{lat74,wan14}, magneto-rotational supernovae (MRSNe, \cite{sym84,nis15,rei21}), and/or magneto-rotational hypernovae/collapsars (MRHNe, \cite{yon21}). Light neutron-capture elements (e.g., Sr) are also produced by electron-capture supernovae (ECSNe, \cite{wan13ecsn}), $\nu$-driven winds \cite{wan13nu}, and rotating-massive stars \cite{fri16,lim18}.

\begin{figure*}
\begin{center}
  \includegraphics[width=0.95\textwidth]{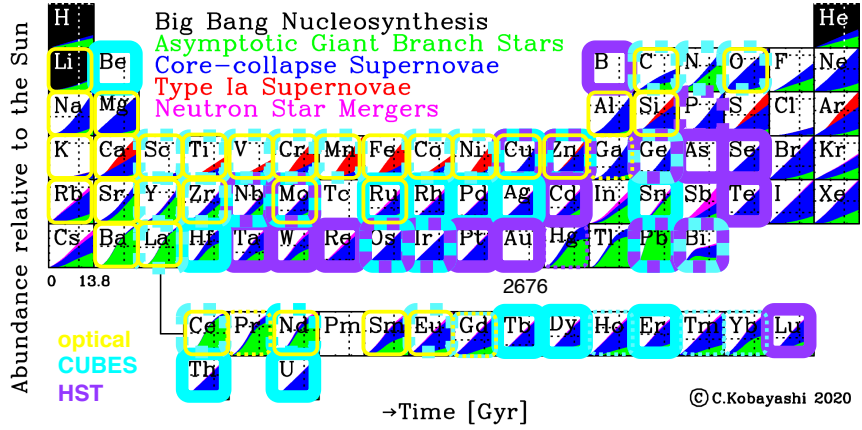}
\caption{The periodic table of elements indicating the key instruments for measuring the stellar abundance, HST (purple), CUBES (cyan), and optical ones such as HERMES (yellow), as well as their origins of elements (colours, see K20 for the details). Thin dotted lines indicates that accessible lines are in the wavelength, but signal-to-noise ratios may not be enough for given spectral resolution. For Ga, only upper limits are obtained with Keck HIRES.}
\label{fig:origin}       
\end{center}
\end{figure*}

The relative contributions depend on the time, and the theoretical prediction from a GCE model for the solar neighborhood was shown in K20.
While most of stable elements are obtained in the solar spectra and/or meteorites (CI chondrites), only limited number of elements are available at low metallicities, some of which require high-resolution IR or UV spectroscopy due to the wavelengths of available absorption lines. In addition, ionized lines or molecular lines are also available in UV, which are important to correct the effects of the departures from local thermodynamic equilibrium (LTE) in stellar abundance analysis.
These elements are highlighted in Figure \ref{fig:origin}.

\section{Evolution of elements}

The time evolutions of elements were shown in K20 in details, and the elements that have new observations are shown in this paper.

\begin{figure}\sidecaption
  \includegraphics[width=0.5\textwidth]{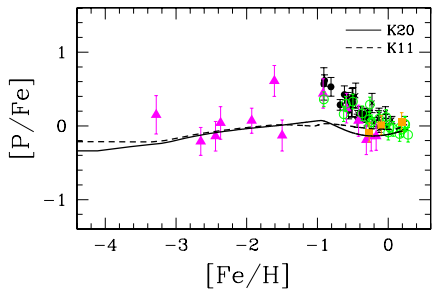}
\caption{The [P/Fe]--[Fe/H] relation comparing with the GCE model predictions from Kobayashi et al. (2020, solid line) and (2011, dashed line). The observational data are taken from \cite{maas17} for crosses; \cite{maas19} for black filled circles; \cite{roe14p} with UV for filled triangles. See K20 for the other data points.
}
\label{fig:p}       
\end{figure}

\begin{figure}\sidecaption
  \includegraphics[width=0.5\textwidth]{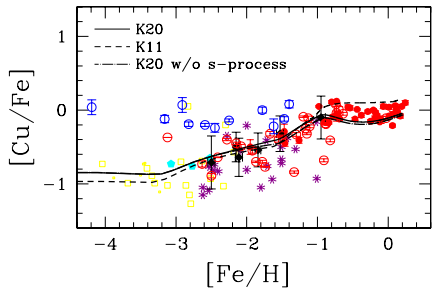}
\caption{The [Cu/Fe]--[Fe/H] relation comparing with UV Cu II abundances from \cite{roe18} for black filled diamonds; and NLTE abundances from \cite{and18} for red open circles, and \cite{shi18} for blue open circles. See K20 for the other data points and model details.}
\label{fig:cu}       
\end{figure}

\begin{figure}\sidecaption
  \includegraphics[width=0.5\textwidth]{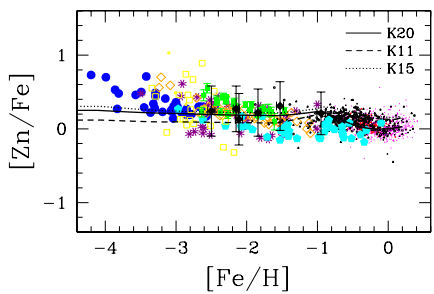}
\caption{The [Zn/Fe]--[Fe/H] relation comparing with UV Zn II abundances from \cite{roe18} for black filled diamonds. See K20 for the other data points and model details.}
\label{fig:zn}       
\end{figure}

Phosphorus ($Z=15$, Fig.~\ref{fig:p}) is an odd-Z element produced by massive stars, and the nucleosynthesis yield becomes smaller for lower metallicity. Thus in the model the [P/Fe] ratio increases by 0.24 dex from [Fe/H] $=-3$ to $-1$, then decreases due to the delayed enrichment from SNe Ia. Although observational data are available with IR for metal-rich stars, HST UV observation is necessary for constraining the evolution and the metallicity dependence. The UV data are in good agreement with the model prediction, and seem to give lower values than the IR data around [Fe/H] $\sim -0.3$. This element could also be a tracer of astrobiology.

Copper ($Z=29$, Fig.~\ref{fig:cu}) is another odd-Z element produced during explosive burning of core-collapse supernovae. The decreasing trend toward low metallicity was known, consistent with the GCE models, but recently, one of the two NLTE results \cite{and18} showed a flat trend. The other NLTE analysis \cite{shi18} and Cu II lines (2126\AA, \cite{roe18}) taken by HST STIS are consistent with the decreasing trend. Since Cu is the only one element that can directly constrain the metallicity dependence of supernova physics, it is very important to obtain the ionized lines.

Zinc ($Z=30$, Fig.~\ref{fig:zn}) is also an important element for constraining supernova physics, namely the explosion energy. The [Zn/Fe] ratio is $\sim 0$ for a wide range of metallicity, which can be reproduced only with the inclusion of hypernovae as in the GCE models. Recently, this is also confirmed by Zn II lines (2139\AA, \cite{roe18}). Around [Fe/H] $\sim -1.3$, although there is no UV data point, the model might slightly overproduce Zn, which is also seen in the recent NLTE analysis \cite{sit22}. This would require a metallicity dependence on the hypernova efficiency. Moreover, in contrast to Cu, [Zn/Fe] shows an increasing trend toward lower metallicity, which indicates that hypernovae may be even more important in the early universe.

\begin{figure*}
\center
  \includegraphics[width=0.95\textwidth]{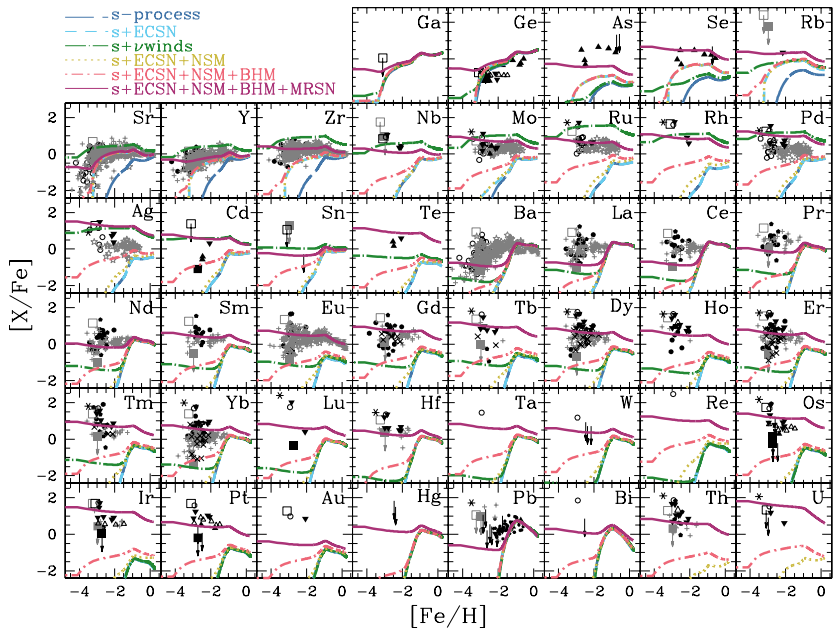}
\caption{The [X/Fe]--[Fe/H] relations for neutron capture elements, comparing to the models in the solar neighborhood, with s-process from AGB stars only (blue long-dashed lines), with s-process and ECSNe (light-blue short-dashed lines), with s-process, ECSNe, and $\nu$-driven winds (green dotted-long-dashed lines), with s-process, ECSNe, and NS-NS mergers (olive dotted lines), with s-process, ECSNe, and NS-NS/NS-BH mergers (orange dotted-short-dashed lines), with s-process, ECSNe, NS-NS/NS-BH mergers, and MRSNe (red solid lines). 
Observational data are updated from K20, including asterisks from \cite{yon21} and open triangles from \cite{cow05}.}
\label{fig:xfe}       
\end{figure*}

Figure \ref{fig:xfe} shows the evolutions of neutron-capture elements. 
As known, AGB stars can produce the first (Sr, Y, Zr), second (Ba), and third (Pb) peak elements, but no heavier elements. It is surprising that ECSNe from a narrow mass range ($\Delta M \sim 0.15-0.2M_\odot$) can enough produce the first peak elements; with the combination of AGB stars, it is possible to reproduce the observational data of most of the elements very well. This means that no other light element primary process (LEPP), such as rotating massive stars, is required. However, the elements from Mo to Ag seem to be overproduced, which could be tested with the UV spectrograph proposed for VLT, CUBES.  Additional production from $\nu$-driven winds lead to further over-production of these elements in the model, but this should be studied with more self-consistent calculations of supernova explosions.

Neutron star mergers (NSMs) can produce lanthanides and actinides, but not enough; the rate is too low and the timescale is too long, according to binary population synthesis. Both nucleosynthesis yields and binary population synthesis of NSMs are uncertain, and for Eu, it is possible to increase the NSM contribution up to 10 times depending on input astrophysical models (Kobayashi et al., in prep.). In any case, it is necessary to have the r-process associated with core-collapse supernovae such as MRSNe. The same conclusion is obtained with other GCE models (see K20 for the references) and more sophisticated chemodynamical simulations \cite{hay19}, as well as from the observational constraints of radioactive nuclei in the solar system \cite{wal21}.

In the GCE model with MRSNe, it is possible to reproduce a plateau at low metallicities for Eu, Pt, and Th, relative to Fe (as shown for Eu in \cite{ces14}'s experiment). However, even with including both MRSNe and NSMs, the predicted Au abundance is more than ten times lower than observed. This underproduction is seen not only for the solar abundance but also for low metallicity stars although the observational data are very limited. UV spectroscopy is necessary for investigating this problem further.

In the GCE models, we adopt the best available nucleosynthesis yields in order to explain the universal r-process pattern: $25M_\odot$ ``b11tw1.00'' 3D yields from \cite{nis15} and $1.3M_\odot$+$1.3M_\odot$ 3D/GR yields from \cite{wan14}.
These yields are sensitive to the electron fraction, which depends on hydrodynamics and $\nu$-processes during explosions. It is necessary to see post-process nucleosynthesis calculations of successful explosion simulations of massive stars.
However, it would not be easy to increase Au yields only since Pt is already in good agreement with the current model. There are uncertainties in some nuclear reaction rates and in the modelling of fission \cite{shibagaki16,vas20}, which might be able to increase Au yields only, without increasing Pt or Ag.
The predicted Th and U abundances are after the long-term decay, to be compared with observations of metal-poor stars, and the current model does not reproduce the Th/U ratio either.

\begin{figure}\sidecaption
  \includegraphics[width=0.5\textwidth]{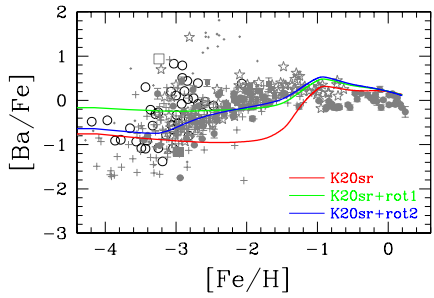}
\caption{The [Ba/Fe]--[Fe/H] relation, comparing with K20 model (red line) and models with rotating massive stars with different rotational velocities (green and blue lines). See K20 for the other data points and model details.}
\label{fig:ba}       
\end{figure}

For s-process elements, there is some underproduction from [Fe/H] $\sim -3$ to $\sim -1$ in the K20 model, which could be improved in chemodynamical simulations. Alternatively, the models with rotating massive stars could better explain the observed average trend of Ba (Fig.~\ref{fig:ba}), La, Ce, Pr, Nd, and Pb, although these models result in overproduction of Sr, Y, and Zr.
In the first model (green line, also plotted in \cite{yon21b}) the same metallicity-dependent distribution of rotational velocity as in \cite{pra18} is used, while in the second model (blue line) 150 and 300 km s$^{-1}$ for HNe and MRHNe, respectively, are assumed.
Note that the nucleosynthesis yields from \cite{lim18} cannot reproduce the observations of many elements, and thus we take only the components of stellar envelopes and winds from their yields and add them to the K20 yields (\cite{kob22}; Kobayashi et al., in prep.).

\section{Conclusions and Discussion}

\begin{table} 
\caption{\label{tab:mrsn}
Relative event numbers, mass ranges of core-collapse supernovae used in our fiducial GCE model, expected remnants (neutron stars or black holes), and necessary conditions for the explosions. See K20 for the details.}
\begin{tabular}{l|ccccc}
\hline
&number & stellar mass [$M_\odot$] & remnant & rotation & magnetic field \\
\hline
ECSN       &47 & $\sim 8.8-9$ &NS& no & no \\
SNII/Ibc   &808& $10-30$ &NS?& no & no \\
failed SN  &19& $30-50$ &BH& no & no \\
HN         &61& $20-50$ &BH?& yes & weak? \\
MRSN/HN       &1& $25-50$ &NS/BH& yes & strong \\
\hline 
\end{tabular} 
\end{table}

Thanks to the long-term collaborations between nuclear and astrophysics, we have good understanding on the origin of elements, except for the elements around Ti and some neutron-capture elements.
Sc, Ti, V, Co, and Zn were shown to be enhanced by multidimensional effects \cite{sne16}, while F, K, Sc, and V were increased by the $\nu$-process in the GCE model of \cite{kob11f}.
Many other elements (e.g., Au, Ag) are accessible only with (near-)UV spectra, the current sample is extremely limited, and thus it has not been possible to discuss the evolution of these elements.
Ionized lines of key elements (e.g., Zn, Cu) and molecular lines (e.g., NH) are also available in UV, which are important to correct the 3D/NLTE effects.
Therefore, UV spectroscopy is needed for further constraining the astronomical sites of the rapid neutron-capture process, the explosion mechanism of core-collapse supernovae, and the effects of stellar rotation, magnetic fields, and binary evolution.

The rates of various types of core-collapse supernovae are summarized in Table 1. In the fiducial model it is assumed that 50\% of stars more than $20M_\odot$ explode as hypernovae to match the observed [Zn/Fe] \cite{kob06}, and that 1.5\% of stars more than $25M_\odot$ produce r-process elements as well in order to reproduce the observed [Eu/Fe] \cite{kob20sr}. This means that the MRSN/HN is rare, one out of $\sim 1000$ supernovae. It is not possible to constrain the fraction between MRSN and MRHN with GCE models, which could be done from the scatters of r-process elements in chemodynamical simulations \cite{hay19} comparing with a larger sample of observations of r-rich extrenely-metal poor stars such as in \cite{yon21b}; these stars are not included enough in Galactic archaeology surveys such as HERMES-GALAH and Gaia-ESO.

The explosion mechanism of BH-forming hypernovae is still unknown, despite a great progress in simulating explosions of NS-forming supernovae \cite{bur21}. The necessary conditions seem to be some rotation of central cores and magnetic fields, but there is no successful explosion simulation \cite{rei21} that successfully produces the observed amount of Fe and actinides \cite{yon21}.
It was also shown that r-process elements can be produced in an accretion disk of a collapsar \cite{sie19}, which is a model proposed for gamma-ray bursts \cite{mac99}. If collapsars explode and produce Fe, that could also be the enrichment source of the star discovered in \cite{yon21}.
Detailed binary evolution could provide a new site, as in \cite{gri22}.

\begin{acknowledgements}
C.K. acknowledges funding from the UK Science and Technology Facility Council (STFC) through grant ST/R000905/1 \& ST/V000632/1, and support from the ``ChETEC'' COST Action (CA16117), supported by COST (European Cooperation in Science and Technology). 
\end{acknowledgements}

\section*{Data Availability Statement}
\vspace*{-1mm}
Chemical evolution model data are available on request at \url{https://star.herts.ac.uk/~chiaki/gce/}. Nucleosynthesis yield data will become public in late 2022.
%
\section*{Conflict of interest}
\vspace*{-1mm}
I have no conflict of interest.



\end{document}